\documentclass[aps,12pt,preprintnumbers,nofootinbib,superscriptaddress]{revtex4}
\usepackage[letterpaper]{geometry}                
\usepackage{graphicx}
\usepackage{amssymb,amsmath}
\def\lag{{\mathcal{L}}}

\def\Dslash{D\hskip-0.65em /}

\begin{document}
\newcount\hour \newcount\minute
\hour=\time \divide \hour by 60
\minute=\time
\count99=\hour \multiply \count99 by -60 \advance \minute by \count99
\newcommand{\mydate}{\ \today \ - \number\hour :00}

\preprint{CALT 68-2656}

\title{Flavor Changing Neutral Currents in the Lee-Wick Standard Model}

\author{Timothy R. Dulaney}
\email{dulaney@theory.caltech.edu}

\author{Mark B. Wise}
\email{wise@theory.caltech.edu}

\affiliation{California Institute of Technology, Pasadena, CA 91125}

\date{\today}                                           

\begin{abstract}

Recently an extension of the standard model (the Lee-Wick standard model) based on ideas of Lee and Wick (LW) was introduced. It does not contain quadratic divergences in the Higgs mass and hence solves the hierarchy puzzle. The LW-standard model contains new heavy LW-resonances at the TeV scale that decay to ordinary particles. In this paper we examine in more detail the flavor structure of the theory. We integrate out the heavy LW-fermions at tree level and find that this induces flavor changing $Z$ boson couplings. However, these flavor changing neutral currents are acceptably small since they are automatically suppressed by small Yukawa couplings.  This is the case even though the theory does not satisfy the principle of minimal flavor violation.  New couplings of the charged $W$ bosons to quarks and leptons are also induced. We also integrate out the LW-Higgs and examine the four-fermion operators induced. 
\end{abstract}

\maketitle
\newpage


\section{Introduction}

The minimal standard model (with dimension five operators to give neutrino's mass) is consistent with almost all experimental observations and possesses many attractive features. For example, it automatically has no renormalizable couplings that violate baryon number and it has a GIM mechanism that suppresses flavor changing neutral currents. Despite the many successes of the standard model,  most elementary particle theorists believe, new physics, beyond the minimal standard model, will emerge at the TeV scale. This is because of the hierarchy problem. In the minimal standard model quadratic divergences in the Higgs mass make it difficult to understand how it can be kept so small compared with the Planck scale.  

Extensions of the standard model that solve the hierarchy puzzle usually contain new degrees of freedom at the TeV scale that can give rise to unacceptably large flavor changing neutral currents.  This happens for example in dynamical symmetry breaking and some versions of low energy supersymmetry.  One way to be sure that new degrees at the TeV scale are not in conflict with experimental constraints on flavor changing neutral currents is to impose the principle of Minimal Flavor Violation (MFV)~\cite{mfv}. In MFV only the standard model Yukawa couplings break the $SU(3)_{Q_L}\times SU(3)_{u_R}\times SU(3)_{d_R}$ quark flavor symmetries.

Recently ideas proposed by Lee and Wick~\cite{Lee:1969fy,Lee:1970iw} were used to extend the standard model so that it does not contain quadratic divergences in the Higgs mass\cite{OGW}. Higher derivative kinetic terms for each of the standard model fields are added that improve the convergence of Feynman diagrams and so there are no quadratically divergent radiative corrections to the Higgs mass. The higher derivative terms give rise to propagators with new poles that are massive resonances. These resonances have wrong-sign kinetic terms which naively cause unacceptable instabilities. Lee and Wick propose altering the energy integrations in the definition of Feynman amplitudes so that the exponential growth does not occur. It appears that this can be done order by order in perturbation theory in a way that preserves unitarity. However, there is acausal behavior caused by the unusual location of poles in the propagators. Physically this acausality is associated with the future boundary condition needed to forbid the exponentially growing modes. As long as the masses and widths of the LW-resonances are large enough, this acausality does not manifest itself on macroscopic scales and is not in conflict with experiment. The possibility of distinguishing LW-resonances from ordinary causal-resonances at the LHC was briefly discussed in Ref.~\cite{rizzo}.  The Lee-Wick photon was searched for experimentally in Ref.~\cite{Busser}.  

The model of Ref.~\cite{OGW} was extended to include right handed neutrinos with masses much above the weak scale.  It was shown that the heavy right handed neutrinos do not give rise to large radiative corrections for the Higgs mass~\cite{neutrino}.  

In~\cite{OGW} minimal flavor violation was imposed to simplify the flavor structure of the theory. In this paper we drop this assumption by allowing for the most general couplings. Integrating out the heavy LW-fermions at tree level we find that there are new flavor changing neutral currents and charged currents, involving both the quark and lepton fields. We calculate the flavor changing $Z$ and $W^\pm$ couplings and argue that they are acceptably small. 

The Lee-Wick standard model introduced in Ref.~\cite{OGW} contains higher derivative terms (\textit{e.g.} the Higgs boson's kinetic term contains two-derivative and four-derivative terms). However there is no principle that forbids the appearance of even higher derivative terms. In fact a Lee-Wick version of the Higgs sector with six-derivative terms was introduced in Ref.~\cite{Kuti}. If even higher derivative terms are present then there are additional LW-resonances. However if the additional LW-resonances occur at substantially higher masses the model presented in Ref.~\cite{OGW} may, for LHC physics, be a reasonable approximation to a theory with even more complicated kinetic terms.

It has been shown that gravity coupled to massless fermions leads to a Lee-Wick theory for gravity~\cite{Tom}.   Recently it was argued that gravitational radiative corrections induce higher derivative terms of the LW-type, leading to a Lee-Wick vector field in the auxiliary field formulation of the theory \cite{Wu}.


\section{The Flavor Sector}

The kinetic terms for the quarks and leptons in the LW-standard model are~\cite{OGW}:
\begin{eqnarray}
&&\lag_\mathrm{hd}^{({\rm kin})} = \overline{\hat Q^i}_L i \hat \Dslash \,\hat Q_L^i + r_Q^{ij} \overline{\hat Q^i}_L i \hat \Dslash \hat \Dslash \hat \Dslash \, \hat Q_L^j+\overline{\hat L^i}_L i \hat \Dslash \,\hat L_L^i + r_L^{ij} \overline{\hat L^i}_L i \hat \Dslash \hat \Dslash \hat \Dslash \, \hat L_L^j +\overline{\hat u^i}_R i \hat \Dslash \,\hat u_R^i \nonumber \\
&&+ r_U^{ij} \overline{\hat u^i}_R i \hat \Dslash \hat \Dslash \hat \Dslash \, \hat u_R^j+\overline{\hat d^i}_R i \hat \Dslash \,\hat d_R^i + r_D^{ij} \overline{\hat d^i}_R i \hat \Dslash \hat \Dslash \hat \Dslash \, \hat d_R^j+\overline{\hat e^i}_R i \hat \Dslash \,\hat e_R^i + r_E^{ij} \overline{\hat e^i}_R i \hat \Dslash \hat \Dslash \hat \Dslash \, \hat e_R^j.
\label{eq:fermionHD}
\end{eqnarray}
Here we have explicity displayed the generation indices ${i,j}$. The $3\times 3$ matrices $r_Q,~r_L,~r_U,~ r_D$ and, $r_E$ are hermitian. In the higher derivative formulation, the fermion Yukawas are
\begin{equation}
{\cal L}^{\rm y} = g_u^{ij} \overline{\hat u^i}_R \hat H \epsilon \hat Q_L^j - g_d^{ij} \overline{\hat d^i}_R \hat H^\dagger \hat Q_L^j - g_e^{ij} \overline{\hat e^i}_R \hat H^\dagger \hat L_L^j + \mathrm{ h.c.} ,
\label{eq:HDyukawa}
\end{equation}
where repeated generation indices in the above equations are summed over ${1,2,3}$. For simplicity we have neglected the neutrino masses.
 
The higher derivative terms can be eliminated by adding massive auxiliary LW-fermion fields: $\tilde Q_L,~\tilde Q_R^\prime,~\tilde L_L,~\tilde L_R^\prime,~\tilde u_R,~\tilde u_L^{\prime},~\tilde d_R,~\tilde d_L^{\prime},~\tilde e_R$ and $\tilde e_L^{\prime}$.  We adopt the notation that the fields that create and destroy the LW resonances are denoted with a ``tilde". Then the ``kinetic" Lagrange density takes the form (suppressing generation indices for simplicity),
\begin{eqnarray}
&&\lag^{({\rm kin})} = \overline{\hat Q}_L i \hat \Dslash \, \hat Q_L +\left( \overline{\tilde Q}_L  M_{\tilde{Q}}\tilde Q_R^\prime + \overline{\tilde Q^\prime}_R  M_{\tilde{Q}} ^{\dagger}\tilde Q_L \right)
+ \overline{\tilde Q}_L i \hat \Dslash \, \hat Q_L + \overline{\hat Q}_L i \hat \Dslash \, \tilde Q_L - \overline{\tilde Q^\prime}_R i \hat \Dslash \, \tilde Q^\prime_R  \nonumber \\
&&+\overline{\hat L}_L i \hat \Dslash \, \hat L_L +\left( \overline{\tilde L}_L  M_{\tilde{L}}\tilde L_R^\prime + \overline{\tilde L^\prime}_R  M_{\tilde{L}} ^{\dagger}\tilde L_L \right)
+ \overline{\tilde L}_L i \hat \Dslash \, \hat L_L + \overline{\hat L}_L i \hat \Dslash \, \tilde L_L - \overline{\tilde L^\prime}_R i \hat \Dslash \, \tilde L^\prime_R \nonumber \\
&&+\overline{\hat u}_R i \hat \Dslash \, \hat u_R +\left( \overline{\tilde u}_R  M_{\tilde{U}}\tilde u_L^\prime + \overline{\tilde u^\prime}_L  M_{\tilde{U}} ^{\dagger}\tilde u_R \right)
+ \overline{\tilde u}_R i \hat \Dslash \, \hat u_R + \overline{\hat u}_R i \hat \Dslash \, \tilde u_R - \overline{\tilde u^\prime}_L i \hat \Dslash \, \tilde u^\prime_L \nonumber \\
&&+\overline{\hat d}_R i \hat \Dslash \, \hat d_R +\left( \overline{\tilde d}_R  M_{\tilde{D}}\tilde d_L^\prime + \overline{\tilde d^\prime}_L  M_{\tilde{D}} ^{\dagger}\tilde d_R \right)
+ \overline{\tilde d}_R i \hat \Dslash \, \hat d_R + \overline{\hat d}_R i \hat \Dslash \, \tilde d_R - \overline{\tilde d^\prime}_L i \hat \Dslash \, \tilde d^\prime_L \nonumber \\
&&+\overline{\hat e}_R i \hat \Dslash \, \hat e_R +\left( \overline{\tilde e}_R  M_{\tilde{E}}\tilde e_L^\prime + \overline{\tilde e^\prime}_L  M_{\tilde{E}} ^{\dagger}\tilde e_R \right)
+ \overline{\tilde e}_R i \hat \Dslash \, \hat e_R + \overline{\hat e}_R i \hat \Dslash \, \tilde e_R - \overline{\tilde e^\prime}_L i \hat \Dslash \, \tilde e^\prime_L,
\end{eqnarray}
where $r_X=\left(M_{\tilde{X}}^{-1}\right)^{\dagger} M_{\tilde{X}}^{-1}$, $X=Q,L,U,D,E$. 

It is convenient to shift the hatted gauge and fermion fields so that their derivative kinetic terms are diagonalized. Using the notation ${\hat{\bf A}}_{\mu}= g_3{\hat A}_{\mu}^B T^B+g_2{\hat W}_{\mu}^aT^a+g_1{\hat B}_{\mu}Y$, it is convenient to make the shifts; ${\hat {\bf A}}_{\mu} ={\bf A}_{\mu}+\tilde {\bf A}_{\mu}$, $\hat Q_L=Q_L-\tilde Q_L$, $\hat L_L=L_L-\tilde L_L$, $\hat u_R=u_R-\tilde u_R$, $\hat d_R=d_R-\tilde d_R$ and $\hat e_R=e_R-\tilde e_R$. Then the fermion kinetic and LW-fermion mass terms become,
\begin{eqnarray}
&&\lag^{({\rm kin})} = \overline{ Q}_L i  \Dslash \,  Q_L +\left( \overline{\tilde Q}_L  M_{\tilde{Q}}\tilde Q_R^\prime + \overline{\tilde Q^\prime}_R  M_{\tilde{Q}} ^{\dagger}\tilde Q_L \right)
- \overline{\tilde Q}_L i \Dslash \, \tilde  Q_L - \overline{\tilde Q^\prime}_R i \Dslash \, \tilde Q^\prime_R  \nonumber \\
&&- \overline{Q_L} \gamma_\mu \tilde{\mathbf{A}}^\mu Q_L
+ \overline{\tilde Q_L} \gamma_\mu \tilde{\mathbf{A}}^\mu \tilde Q_L + \overline{\tilde Q^\prime_R} \gamma_\mu \tilde{\mathbf{A}}^\mu \tilde Q_R^\prime \nonumber \\
&&+\overline{ L}_L i  \Dslash   L_L +\left( \overline{\tilde L}_L  M_{\tilde{L}}\tilde L_R^\prime + \overline{\tilde L^\prime}_R  M_{\tilde{L}} ^{\dagger}\tilde L_L \right)
- \overline{\tilde L}_L i  \Dslash \, \tilde L_L  - \overline{\tilde L^\prime}_R i  \Dslash \, \tilde L^\prime_R \nonumber \\
&&- \overline{L_L} \gamma_\mu \tilde{\mathbf{A}}^\mu L_L
+ \overline{\tilde L_L} \gamma_\mu \tilde{\mathbf{A}}^\mu \tilde L_L + \overline{\tilde L^\prime_R} \gamma_\mu \tilde{\mathbf{A}}^\mu \tilde L_R^\prime \nonumber \\
&&+\overline{ u}_R i  \Dslash  u_R +\left( \overline{\tilde u}_R  M_{\tilde{U}}\tilde u_L^\prime + \overline{\tilde u^\prime}_L  M_{\tilde{U}} ^{\dagger}\tilde u_R \right)
- \overline{\tilde u}_R i  \Dslash \, \tilde  u_R  - \overline{\tilde u^\prime}_L i  \Dslash \, \tilde u^\prime_L \nonumber \\
&&- \overline{u_R} \gamma_\mu \tilde{\mathbf{A}}^\mu u_R
+ \overline{\tilde u_R} \gamma_\mu \tilde{\mathbf{A}}^\mu \tilde u_R + \overline{\tilde u^\prime_L} \gamma_\mu \tilde{\mathbf{A}}^\mu \tilde u_L^\prime \nonumber \\
&&+\overline{ d}_R i  \Dslash \,  d_R +\left( \overline{\tilde d}_R  M_{\tilde{D}}\tilde d_L^\prime + \overline{\tilde d^\prime}_L  M_{\tilde{D}} ^{\dagger}\tilde d_R \right)
- \overline{\tilde d}_R i \Dslash \, \tilde  d_R - \overline{\tilde d^\prime}_L i  \Dslash \, \tilde d^\prime_L \nonumber \\
&&- \overline{d_R} \gamma_\mu \tilde{\mathbf{A}}^\mu d_R
+ \overline{\tilde d_R} \gamma_\mu \tilde{\mathbf{A}}^\mu \tilde d_R + \overline{\tilde d^\prime_L} \gamma_\mu \tilde{\mathbf{A}}^\mu \tilde d_L^\prime \nonumber \\
&&+\overline{ e}_R i  \Dslash \, e_R +\left( \overline{\tilde e}_R  M_{\tilde{E}}\tilde e_L^\prime + \overline{\tilde e^\prime}_L  M_{\tilde{E}} ^{\dagger}\tilde e_R \right)
- \overline{\tilde e}_R i \Dslash \, \tilde e_R  - \overline{\tilde e^\prime}_L i \Dslash \, \tilde e^\prime_L \nonumber \\
&&- \overline{e_R} \gamma_\mu \tilde{\mathbf{A}}^\mu e_R
+ \overline{\tilde e_R} \gamma_\mu \tilde{\mathbf{A}}^\mu \tilde e_R + \overline{\tilde e^\prime_L} \gamma_\mu \tilde{\mathbf{A}}^\mu \tilde e_L^\prime.
\end{eqnarray}
With these redefinitions and the change $\hat H=H-\tilde H$ the Yukawa interactions become,
\begin{eqnarray}
&& {\cal L}^{({\rm y})} = (\overline{u}_R - \overline{\tilde u}_R) g_u (H - \tilde H) \epsilon (Q_L - \tilde Q_L) -  (\overline{d}_R - \overline{\tilde d}_R )g_d (H^\dagger - \tilde H^\dagger) (Q_L - \tilde Q_L) \nonumber \\
&& - (\overline{e}_R - \overline{\tilde e}_R) g_e (H^\dagger - \tilde H^\dagger) (L_L - \tilde L_L) + \mathrm{ h.c.} ,
\label{eq:LWyukawa}
\end{eqnarray}
where the generation indices have been suppressed.
The LW-field mass matrices are diagonalized by the unitary transformations,
\begin{equation}
\tilde Q_L \rightarrow Y(\tilde Q_L)\tilde Q_L,~~~~\tilde Q_R^\prime \rightarrow Y(\tilde Q_R^\prime)\tilde Q_R^\prime,~~~~Y(\tilde Q_L)^{\dagger}M_{\tilde{Q}} Y(\tilde Q_R^\prime)=M_{\tilde{Q}}^{\rm (diag)},
\end{equation}
\begin{equation}
\tilde L_L \rightarrow Y(\tilde L_L)\tilde L_L,~~~~\tilde L_R^\prime \rightarrow Y(\tilde L_R^\prime)\tilde L_R^\prime,~~~~Y(\tilde L_L)^{\dagger}M_{\tilde{L}} Y(\tilde L_R^\prime)=M_{\tilde{L}}^{\rm (diag)},
\end{equation}
\begin{equation}
\tilde u_R \rightarrow Y(\tilde u_R)\tilde u_R,~~~~\tilde u_L^\prime \rightarrow Y(\tilde u_L^\prime)\tilde u_L^\prime,~~~~Y(\tilde u_R)^{\dagger}M_{\tilde{U}} Y(\tilde u_L^\prime)=M_{\tilde{U}}^{\rm (diag)},
\end{equation}
\begin{equation}
\tilde d_R \rightarrow Y(\tilde d_R)\tilde d_R,~~~~\tilde d_L^\prime \rightarrow Y(\tilde d_L^\prime)\tilde d_L^\prime,~~~~Y(\tilde d_R)^{\dagger}M_{\tilde{D}} Y(\tilde d_L^\prime)=M_{\tilde{D}}^{\rm (diag)}
\end{equation}
and
\begin{equation}
\tilde e_R \rightarrow Y(\tilde e_R)\tilde e_R,~~~~\tilde e_L^\prime \rightarrow Y(\tilde e_L^\prime)\tilde e_L^\prime,~~~~Y(\tilde e_R)^{\dagger}M_{\tilde{E}} Y(\tilde d_L^\prime)=M_{\tilde{E}}^{\rm (diag)}.
\end{equation}
These transformations do not, via the kinetic terms, introduce any gauge boson couplings to the heavy LW-fermions that change generation.  The coupling matrices from the higher derivative theory can be written in terms of these transformations, for example,
\begin{equation}
r_Q = Y(\tilde Q_L) \left(\frac{1}{M_{\tilde{Q}}^{\rm (diag)}}\right)^2 Y(\tilde Q_L)^{\dagger} ~~{\rm and}~~ r_E = Y(\tilde e_R) \left(\frac{1}{M_{\tilde{E}}^{\rm (diag)}}\right)^2 Y(\tilde e_R)^{\dagger}.
\end{equation}
The Yukawas take the form,
\begin{eqnarray}
&& {\cal L}^{({\rm y})} = (\overline{u}_R - \overline{\tilde u}_RY(\tilde u_R)^{\dagger}) g_u (H - \tilde H) \epsilon (Q_L - Y(\tilde Q_L)\tilde Q_L) - \nonumber \\ 
&&  (\overline{d}_R - \overline{\tilde d}_R Y(\tilde d_R)^{\dagger})g_d (H^\dagger - \tilde H^\dagger) (Q_L - Y(\tilde Q_L)\tilde Q_L) \nonumber \\
&&- (\overline{e}_R - \overline{\tilde e}_RY(\tilde e_R)^{\dagger}) g_e (H^\dagger - \tilde H^\dagger) (L_L - Y(\tilde L_L)\tilde L_L) + \mathrm{ h.c.}~.
\end{eqnarray}
This completes this formulation of the version of the theory with auxiliary fields and without higher derivatives.


\section{New W and Z Couplings}

When the higher derivative terms are treated as a perturbation one can apply the equations of motion for the quark and the lepton fields to get,
\begin{eqnarray}
\label{hatops}
&&{\cal L}^{({\rm eff})}= +\overline{\hat{u}}_R\left(g_u r_Q g_u^{\dagger}\right) \hat{H}^T \epsilon i  \hat{\Dslash} \, (\epsilon \hat{H}^* \hat{u}_R) - \overline{\hat{d}}_R\left(g_d  r_Q g_d^{\dagger}\right) \hat{H}^{\dagger}i  \hat{\Dslash} \, (\hat{H} \hat{d}_R) \nonumber \\
&&-\overline{\hat{e}}_R\left(g_e r_L g_e^{\dagger} \right)\hat{H}^{\dagger}i  \hat{\Dslash} \, (\hat{H} \hat{e}_R) + [\overline{\hat{Q}}_L \epsilon \hat{H}^*]\left(g_u^{\dagger} r_U g_u \right)i \hat{\Dslash} \, [\hat{H}^T \epsilon \hat{Q}_L] \nonumber \\
&&- [\overline{\hat{Q}}_L \hat{H}]\left(g_d^{\dagger} r_D g_d \right)i \hat{\Dslash} \, [\hat{H} ^{\dagger}\hat{Q}_L] - [\overline{\hat{L}}_L \hat{H}]\left(g_e^{\dagger} r_E g_e \right)i \hat{\Dslash} \, [\hat{H} ^{\dagger} \hat{L}_L] \nonumber \\
&&-\left\{\overline{\hat{d}}_R\left(g_d r_Q g_u^{\dagger}\right)\hat{H}^{\dagger} i \hat{\Dslash} \, [\epsilon \hat{H}^* \hat{u}_R] +{\rm h.c.} \right\}.
\end{eqnarray}
As we shall see, this yields new flavor changing $Z$ boson couplings and new charged $W^\pm$ boson couplings.  

\begin{figure}[!htp]
\centering
\includegraphics[trim= 2in 9in 1in 0.8in ]{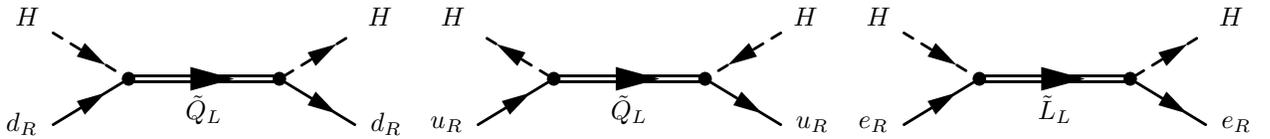}
\caption{Diagrams used to integrate out the heavy LW-fermions resulting in flavor changing neutral currents (FCNCs) involving the $Z$ boson.}\label{FCNC}
\end{figure}

\begin{figure}[!htp]
\centering
\includegraphics[trim= 1in 9in 0.5in 0.8in ]{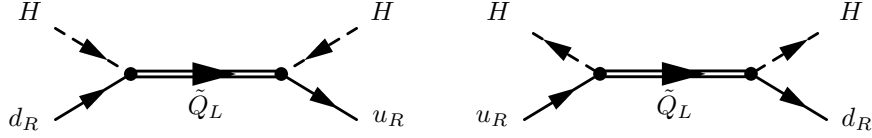}
\caption{Diagrams used to integrate out the heavy LW-fermions resulting in additional flavor changing charged currents (FCCCs) involving the $W^\pm$ bosons.}\label{FCCC}
\end{figure}

It is instructive to also derive the above result from the version of the theory with auxiliary fields by integrating out the LW-fermions at tree-level.  This gives rise to an effective theory where the LW-fermions are absent.  In figure (\ref{FCNC},~\ref{FCCC}) we have a few examples of the tree-level diagrams considered when integrating out the Lee-Wick fermions, represented by double-lines, to derive the flavor changing (neutral, charged) currents.  

\newpage
The effective Lagrangian is determined by calculating the Feynman diagrams in Figs.~\ref{FCNC} and \ref{FCCC}.   Gauge invariance is then used to convert ordinary derivatives to covariant derivatives.  Alternatively, the LW-fermions can be integrated out using their equations of motion at low-energy.  These approaches result in an effective theory with the following dimension-six operators with unusual $W$ and $Z$ couplings to standard model fermions,
\begin{eqnarray}
\label{ops}
&&{\cal L}^{({\rm eff})}= +\overline{u}_R\left(g_u r_Q g_u^{\dagger}\right)H^T \epsilon i  \Dslash \, (\epsilon H^* u_R) - \overline{d}_R\left(g_d  r_Q g_d^{\dagger}\right)H^{\dagger}i  \Dslash \, (H d_R) \nonumber \\
&&-\overline{e}_R\left(g_e r_L g_e^{\dagger} \right)H^{\dagger}i  \Dslash \, (H e_R) + [{\overline Q}_L \epsilon H^*]\left(g_u^{\dagger} r_U g_u \right)i\Dslash \, [H^T \epsilon Q_L] \nonumber \\
&&- [{\overline Q}_L H]\left(g_d^{\dagger} r_D g_d \right)i\Dslash \, [H ^{\dagger}Q_L] - [{\overline L}_L H]\left(g_e^{\dagger} r_E g_e \right)i\Dslash \, [H ^{\dagger} L_L] \nonumber \\
&&-\left\{\overline{d}_R\left(g_d r_Q g_u^{\dagger}\right)H^{\dagger} i \Dslash \, [\epsilon H^* u_R] +{\rm h.c.}. \right\}
\end{eqnarray}
In the above expression, when there is possible ambiguity about how the $SU(2)$ singlets are formed, we have used square brackets to denote the contraction of $SU(2)$ indices.  The covariant derivative is understood to act on all fields to the right of it.  When the Higgs doublet gets its vacuum expectation value the standard model fermions get their masses in the usual way,
\begin{equation}
m_u=g_u v /{\sqrt 2},~~m_d=g_d v/{\sqrt 2},~~m_e=v g_e/{\sqrt 2}.
\end{equation}
 These mass matrices are diagonalized by the following transformations,
\begin{equation}
u_L \rightarrow {\cal U}(u,L)u_L,~~ u_R \rightarrow {\cal U}(u,R)u_R , ~~ {\cal U}(u,R)^{\dagger} m_u {\cal U}(u,L)=m_u^{\rm diag},
\end{equation}
\begin{equation}
d_L \rightarrow {\cal U}(d,L)d_L,~~ d_R \rightarrow {\cal U}(d,R)d_R , ~~ {\cal U}(d,R)^{\dagger} m_d {\cal U}(d,L)=m_d^{\rm diag},
\end{equation}
and
\begin{equation}
e_L \rightarrow {\cal U}(e,L)e_L,~~ e_R \rightarrow {\cal U}(e,R)e_R , ~~ {\cal U}(e,R)^{\dagger} m_e {\cal U}(e,L)=m_e^{\rm diag}.
\end{equation}
Giving the Higgs its vacuum expectation value and going to the standard model quark and lepton mass matrices we find that Eq.~(\ref{ops}) gives rise to the following new Z-boson couplings,
\begin{eqnarray}
\label{zee}
&&\Delta{\cal L}_Z= {\sqrt{g_1^2+g_2^2}}Z_{\mu}\left[\overline{u}_R m_u^{\rm diag}\left( {\cal U}(u,L)^{\dagger} r_Q \, {\cal U}(u,L)\right)m_u^{\rm diag} \gamma^{\mu}u_R \right. \nonumber \\
&&+\overline{d}_R m_d^{\rm diag}\left( {\cal U}(d,L)^{\dagger} r_Q \, {\cal U}(d,L)\right)m_d^{\rm diag} \gamma^{\mu}d_R + \overline{e}_R m_e^{\rm diag}\left( {\cal U}(e,L)^{\dagger} r_L \, {\cal U}(e,L)\right)m_e^{\rm diag} \gamma^{\mu}e_R \nonumber \\
&&+\overline{u}_L m_u^{\rm diag}\left( {\cal U}(u,R)^{\dagger} r_U \, {\cal U}(u,R)\right)m_u^{\rm diag} \gamma^{\mu}u_L +\overline{d}_L m_d^{\rm diag}\left( {\cal U}(d,R)^{\dagger} r_D \, {\cal U}(d,R)\right)m_d^{\rm diag} \gamma^{\mu}d_L \nonumber \\
&& + \left. \overline{e}_L m_e^{\rm diag}\left( {\cal U}(e,R)^{\dagger} r_E \, {\cal U}(e,R)\right)m_e^{\rm diag} \gamma^{\mu}e_L \right],
\end{eqnarray}
and the following new $W$ boson couplings
\begin{eqnarray}
&& \Delta{\cal L}_W= {g_2 \over \sqrt 2}W_{\mu}^- \left[\overline{d}_R m_d^{\rm diag} \left( {\cal U}(d,L)^{\dagger}r_Q \, {\cal U}(u,L)\right)m_u^{\rm diag} \gamma^{\mu}u_R \right. \nonumber \\
&& + \frac{1}{2} \overline{d}_L V^\dagger m_u^{\rm diag} \left( {\cal U}(u,R)^{\dagger}r_U \, {\cal U}(u,R)\right)m_u^{\rm diag} \gamma^{\mu}u_L \nonumber \\
&& +  \frac{1}{2} \overline{d}_L m_d^{\rm diag} \left( {\cal U}(d,R)^{\dagger}r_D \, {\cal U}(d,R)\right)m_d^{\rm diag} \gamma^{\mu} V^\dagger  u_L \nonumber \\
&& \left. +  \frac{1}{2}  \overline{e}_L m_e^{\rm diag} \left( {\cal U}(e,R)^{\dagger}r_E \, {\cal U}(e,R)\right)m_e^{\rm diag} \gamma^{\mu} \nu_L \right] + {\rm h.c.},
\end{eqnarray}
where $V={\cal U}(u,L)^{\dagger}{\cal U}(d,L)$ is the usual standard model CKM matrix.  Note that in each case of the flavor changing neutral currents (FCNCs), involving coupling to the $Z$ boson, the matrix in flavor space is simply a unitary rotation of the higher derivative couplings.  The last of the FCNCs is actually a lepton-family violating current.  This leads to, for example, a tree-level vertex for $\mu^\pm \rightarrow Ze^\pm$.  The same can be said for the new flavor changing charged currents (FCCC), involving coupling of the $W^\pm$ bosons to the quarks.  In this case, the matrix in flavor space is a unitary rotation of the higher derivative coupling multiplied by the standard model CKM matrix.  

Minimal flavor violation (MFV) corresponds to the case where the heavy LW-fermion mass matrices are proportional to the identity matrix.   Thus, in MFV, the flavor changing couplings of the $Z$ vanish, the unusual right-handed quark couplings to the charged $W$-bosons are proportional to the mixing angles in the CKM matrix $V$ and the lepton-family violating interaction vanishes.

The flavor changing $Z$ couplings in Eq.~(\ref{zee}) are very small even if the LW-fermion mass matrices are not proportional to the  identity. This is because of the standard model quark and lepton mass factors. For example if the LW-fermions are of order the TeV scale then Eq.~(\ref{zee}) gives a $\mu \rightarrow 3e$ rate of order,
\begin{equation}
{\Gamma(\mu \rightarrow 3e) \over \Gamma (\mu \rightarrow e\bar \nu_e \nu_{\mu})}\sim \left({m_em_{\mu} \over 1{\rm TeV}^2}\right)^2 \sim 10^{-21}.
\end{equation}
The effects can be larger in cases where there is a standard model contribution to interfere with. For example, the change in the rate for $b \rightarrow s \nu \bar \nu$ is of order,
\begin{equation}
{\Delta\Gamma(b \rightarrow s \nu \bar \nu) \over \Gamma (b \rightarrow s \nu \bar \nu)}\sim \left({4\pi {\rm sin}^2\theta_W \over \alpha |V_{cb}|}\right)\left(m_b m_s \over 1{\rm TeV}^2\right)\sim 10^{-6}.
\end{equation}
The new couplings of the charged $W$-bosons to the right-handed quark fields are similarly suppressed.

The new couplings of the $W^\pm$ bosons to the left-handed quark and lepton fields are not necessarily suppressed by the ratio of the tree-level standard model fermion masses participating in the interaction to the TeV scale masses of LW-resonances.  However, these couplings are still small.  For example, inclusion of these currents results in a change of the effective, or observed, matrix element $V_{cb}$ of,
\begin{equation}
\Delta V_{cb} \sim \frac{m_c m_t}{(1 TeV)^2} \sim 10^{-4}.
\end{equation}
Similarly there are small corrections to the other effective CKM matrix elements.  Note that the effective CKM matrix is not unitary.  


\section{Four-Fermion Operators From Inegrating out the LW-Higgs}
For completeness in deriving the low energy effective field theory and to give the operators that contribute at tree-level to interesting processes, we now integrate out the LW-partner of the Higgs boson at tree-level. Note that in this section we are using the notation for the quark doublet $Q'_L = (u_L, V d_L)$ and  $Q''_L = (V^\dagger u_L, d_L)$, where V is the standard model CKM Matrix.  The resulting effective field theory in which the LW-Higgs is absent gives four-fermion operators, of dimension-six, suppressed by the mass of the LW-Higgs resonance ($M_{\tilde{H}}$),
\begin{eqnarray}
\label{Hops}
&& {\cal L}^{({\rm eff})}= -\frac{2}{M^2_{\tilde{H}}} \left[ -[{\overline Q'}_L \left( \frac{m_u^{\rm diag}}{v} \right)  u_R] [{\overline u}_R \left(\frac{m_u^{\rm diag}}{v}\right) Q'_L]  \right. \nonumber \\
&& +[{\overline Q''}_L \left(\frac{m_d^{\rm diag}}{v}\right)  d_R] [{\overline d}_R \left(\frac{m_d^{\rm diag}}{v}\right) Q''_L]  + [{\overline L}_L \left(\frac{m_e^{\rm diag}}{v}\right)  e_R] [{\overline e}_R \left(\frac{m_e^{\rm diag}}{v}\right) L_L]   \nonumber \\
&&+ \left( [{\overline Q''}_L \left(\frac{m_d^{\rm diag}}{v}\right)  d_R] [{\overline e}_R \left(\frac{m_e^{\rm diag}}{v}\right) L_L] + {\rm h.c}\right) \nonumber \\
&&  + \left(  [{\overline u}_R \left( \frac{m_u^{\rm diag}}{v}\right) {Q'_L}^T]\epsilon [{\overline e}_R \left(\frac{m_e^{\rm diag}}{v}\right) L_L]  + {\rm h.c}\right)  \nonumber \\
&& \left. +  \left([{\overline u}_R \left(\frac{m_u^{\rm diag}}{v}\right) {Q'_L}^T] \epsilon [{\overline d}_R \left(\frac{m_d^{\rm diag}}{v}\right) Q''_L] + {\rm h.c}\right) \right].
\end{eqnarray}
In the equation above we use square brackets to denote Lorentz scalars and to make clear how the generation indices are being summed.  There is no ambiguity about how $SU(2)$ singlets are formed.  The effective Lagrangian in Eq.~(\ref{Hops}) gives very small contributions to standard model processes (\textit{e.g.} $B-\bar{B}$-mixing, $K \rightarrow \pi e^+ e^-$, \textit{etc.}).


\section{Conclusion}
In this paper we have considered the Lee-Wick (LW) extension of the minimal Standard Model proposed by \cite{OGW}.  Rather than imposing the principle of Minimal Flavor Violation (MFV), forcing all LW-fermions in the same representation of the gauge group to have the same mass, we explored the most general flavor structure of the theory by imposing no constraints (beyond hermiticity) on the higher derivative couplings.  We present the resulting auxiliary field formulation of the theory, involving heavy LW auxiliary fields and showed explicitly the relation to the higher derivative theory.  From the auxiliary field formulation involving LW-fields, we derive a low energy effective field theory by integrating out the heavy auxiliary fields at tree-level.  We show that new flavor changing neutral currents (FCNCs) and charged currents (FCCCs) are present in the theory. 

One would naively expect that these FCCCs and FCNCs would lead to unacceptably large mixings (at odds with experimental results), but we show that these mixings are appropriately suppressed if the LW auxiliary fields are at the TeV scale.  This suppression is a general feature of the theory even though the theory does not satisfy the principle of MFV.  We show further that when the principle of MFV is applied that the FCNCs vanish and the FCCCs simply reduce to small additional mixings of the usual Standard Model CKM type, involving both left and right-handed quark fields (recovering the result stated in \cite{OGW} where MFV was applied). 

For completeness, we also found the four-fermion operators resulting from integrating out the LW-Higgs.   They are all suppressed by the LW-Higgs mass, and give very small contributions to standard model processes.  

This paper highlights an attractive feature of the Lee-Wick extension of the minimal standard model.  This extension does not require the \textit{ad hoc} application of the principle of Minimal Flavor Violation to adequately suppress additional flavor changing currents in the low energy effective field theory.  

We thank B. Grinstein for helpful comments on the derivation of Eq.~(\ref{hatops}).  We also thank him and D. O'Connell for encouraging remarks on a preliminary version of this manuscript.  This work was supported in part by DOE grant number DE-FG03-92ER40701.




\begin{thebibliography}{99}
\bibitem{mfv}
   R.~S.~Chivukula and H.~Georgi,
   Phys.\ Lett.\  B {\bf 188}, 99 (1987);
   L.~J.~Hall and L.~Randall,
   Phys.\ Rev.\ Lett.\ {\bf 65}, 2939 (1990);
  G.~D'Ambrosio, G.~F.~Giudice, G.~Isidori and A.~Strumia,
  Nucl.\ Phys.\  B {\bf 645}, 155 (2002)
  [arXiv:hep-ph/0207036].
  
\bibitem{Lee:1969fy}
  T.~D.~Lee and G.~C.~Wick,
  Nucl.\ Phys.\  B {\bf 9}, 209 (1969).

\bibitem{Lee:1970iw}
  T.~D.~Lee and G.~C.~Wick,
  Phys.\ Rev.\  D {\bf 2}, 1033 (1970).
\bibitem{OGW}
B.~Grinstein, D.~O'Connell and M.~B.~Wise,
  arXiv:0704.1845 [hep-ph].
\bibitem{rizzo}
 T.~G.~Rizzo,
  JHEP {\bf 0706}, 070 (2007)
  [arXiv:0704.3458 [hep-ph]].

\bibitem{Busser}
  B\"usser, F.W., et al. 
  Phys.\ Lett.\ B {\bf 48}, 377 (1974).

\bibitem{neutrino}
J.~R.~Espinosa, B.~Grinstein, D.~O'Connell and M.~B.~Wise,
  arXiv:0705.1188 [hep-ph].

\bibitem{Kuti}
  K.~Jansen, J.~Kuti and C.~Liu,
  Phys.\ Lett.\  B {\bf 309}, 119 (1993)
  [arXiv:hep-lat/9305003];
  K.~Jansen, J.~Kuti and C.~Liu,
  Phys.\ Lett.\  B {\bf 309}, 127 (1993)
  [arXiv:hep-lat/9305004].

\bibitem{Tom}
  E. Tomboulis,
  Phys.\ Lett.\ B {\bf 70}, 361 (1977); 
  E. Tomboulis,
  Phys.\ Lett.\ B {\bf 97}, 77 (1980).
  
\bibitem{Wu}
  F.~ Wu and M.~Zhong,
  arXiv:0705.3287 [hep-ph].


\end{thebibliography}
\end{document}